# A 128-channel picoammeter system and its application on charged particle beam current distribution measurements

Deyang Yu,[a)] Junliang Liu, Yingli Xue, Mingwu Zhang, Xiaohong Cai, Jianjun Hu, and Jinmei Dong

*Institute of Modern Physics, Chinese Academy of Sciences, Lanzhou 730000, China*

Xin Li

*School of Nuclear Science and Technology, Lanzhou University, Lanzhou 730000, China*

A 128-channel picoammeter system is constructed based on instrumentation amplifiers. Taking the advantages of high electric potential and narrow bandwidth in DC energetic charged beam measurements, current resolution better than 5 fA can be achieved. Two 128-channel Faraday cup arrays are built, and are employed for ion and electron beam current distribution measurements. Tests with 60 keV $O^{3+}$ ions and 2 keV electrons show that it can provide exact boundaries when a positive charged particle beam current distribution is measured.

## I. INTRODUCTION

The absolute current distribution is usually the most desirable parameter in many experiments when a charged particle beam is involved. The total beam intensity can be precisely measured by a Faraday cup [1-6] or a beam current density meter.[7] Meanwhile, the profile can be obtained by various beam monitors, such as a fixed multi-wire beam profile monitor,[8] a single-wire scanner beam profile monitor,[9-16] a scintillation screen,[17,18] a residual gas monitor.[19-22] Recently, the Faraday cup array (FCA) technique by directly measuring absolute current distribution are developed both in one dimension [23] and two dimensions,[24] with great advantage of convenience. In this technique, a many-channel picoammeter (up to 128 channels in the present application) is necessary. However, to our best knowledge, nowadays there are only a few channels available in commercial picoammeters, such as a Keithley Model 6482 dual channel picoammeter [25] or a CAENels Model AH401D [26] or AH501D [27] 4-channel picoammeter, which are far from requirement. As a result, special electronic schemes [23,24,28,29] are developed to deal with all the independent channels of a FCA. It requires a very high sensitivity of the electronics, when taking into account that a beam current of picoamperes is distributed on many independent channels. In practice, the sensitivity is limited by environment interference, leak currents from the suppressing voltage to the electrodes, noise of the electronics and escape of the secondary electrons. Current resolutions of 10 pA and 50 pA were reported in previous works, which correspond to beam fluxes of 100 pA/cm$^2$ in one dimensional [23] and 200 nA/cm$^2$ in two dimensional measurement, [24] respectively.

In this paper, we present a one-dimensional absolute beam current distribution measurement system which is consisted of a 128-channel picoammeter in combination with a 128-channel FCA. Current resolution of 5 fA for each channel and position resolution of 0.3 mm/channel are achieved.

## II. THE PICOAMMETER SYSTEM

The main challenge to the electronics is to ensure a very high sensitivity of many independent channels in compact space. However, different from other applications, [30-34] two significant advantages can be utilized to strongly suppress the noises and therefore enhance the sensitivity in measurement of an energetic charged particle beam. First, the electrodes of a FCA which collect the beam are almost ideal current sources, with high electric potentials and infinite internal resistances. As a result, a high input resistance of the picoammeter up to several GΩ is acceptable. Considering an electrode which accepts 1 nA beam current, an input resistance of 1 GΩ will push its electric potential to 1 V, which is negligible comparing to the beam potential. Second, a quasi-direct beam current measurement requires only a very narrow bandwidth down to below 1 Hz.

Instead of a traditional transresistance amplifier based on an operational amplifier, [23,34-42] a current-to-voltage converter based on an instrumentation amplifier is implemented in a double-sided printed circuit board (PCB) for each channel, i.e., the daughter board. The circuit diagram and the photo of the daughter board are shown in Fig. 1. In particular, a very narrow bandwidth low-pass filter, consisted of the resistor $R_F$ and the capacitor $C_F$, is inserted before the current-to-voltage conversion resistor $R_C$ to suppress the noises and interference. The response of

---

the input stage to an induced noise $v(\omega)$ is

$$H_v(\omega) = \frac{1}{1+R_F/R_C+j\omega R_F C_F}, \quad (1)$$

where $j = \sqrt{-1}$ is the imaginary unit. When a current signal $i(\omega)$ is input, the output voltage of this input stage is $i(\omega)R_C/(1 + j\omega R_C C_F)$. However, if $R_F$ and $C_F$ do not exist, the input current is equivalent to an input voltage of $i(\omega)R_C$. Therefore the effective response to a current $i(\omega)$ is

$$H_i(\omega) = \frac{1}{1+j\omega R_C C_F}. \quad (2)$$

These equations show that the filter's response to a current signal or an induced noise can be controlled independently, and scarcely influences the quasi-direct component of the beam current signal. In particular, $R_F$ has no effect on conversion of the current signal. In the present application, the noise mainly comes from the electric induction by the 50 Hz AC power supply in our circumstance. In contrast, the beam is a quasi-direct current varying slower than 0.1 Hz. The resistor $R_F = 1$ GΩ and the capacitor $C_F = 1$ nF are employed, while the resistor $R_C$ is designed to be selectable from 1 GΩ, 200 MΩ, 50 MΩ and 10 MΩ on the daughter board, to cover a wider measurement range. For example, when the current is a few pA/channel, $R_C = 1$ GΩ should be selected, and the cutoff frequency to noise and to signal is about 0.08 Hz and 0.16 Hz, respectively. While a stronger beam of a few nA/channel is measured, $R_C = 200$ MΩ can be selected, and the cutoff frequency to noise and to signal is about 0.15 Hz and 0.8 Hz, respectively.

The converted voltage signal across the resistor $R_C$, is amplified by a following INA116 of Texas Instruments.[43] It is a complete monolithic FET-input instrumentation amplifier with a common-mode rejection ratio of 84 dB, a bias current of 3 fA and an input resistance of $10^{15}$ Ω. As a consequence the shunt current is negligible considering the converter resistor $R_C$. In order to inhibit the leakage currents from the power supply to the non-inverting and the inverting input stages, they are guarded by ground rings cooperating with guard pins, respectively. Take the advantage of the very narrow bandwidth, the guarding pins and the reference pin of the chip are grounded together, as shown in the insert photo of Fig. 1. The voltage gain of INA116 itself is controlled by a single resistor $R_G$, which is also selectable on the daughter board from 100 Ω, 511 Ω, 2.61 kΩ and 12.4 kΩ, respectively. The voltage gain of INA116 is $1 + 50\text{ k}\Omega/R_G$, and therefore the transfer gain of the present current-to-voltage converter is

$$G[\text{V/A}] = \left(1 + \frac{50\text{ k}\Omega}{R_G}\right)R_C. \quad (3)$$

Resistors of $R_C$ and $R_G$ with high stability and low temperature coefficient are employed to guarantee the gain stability. In order to further inhibit the inductive noise on the cable between the amplifier output and the acquisition card, another low-pass RC circuit is inserted following the output of the INA116, in consideration of the quasi-direct feature of the signal and the high input resistance of the acquisition card. It is composed of the resistor $R_O$ and the capacitor $C_O$, as shown in Fig. 1. In the present version $R_O = 1$ kΩ and $C_O = 10$ μF is selected. We note that the resistor $R_O$ also acting as an output protection to the INA116, and the capacitor $C_O$ actually provides a ground route for the inductive noise. In addition, the ±12 V power supply of the chip are protected by diodes and filtered by 10 μF capacitors.

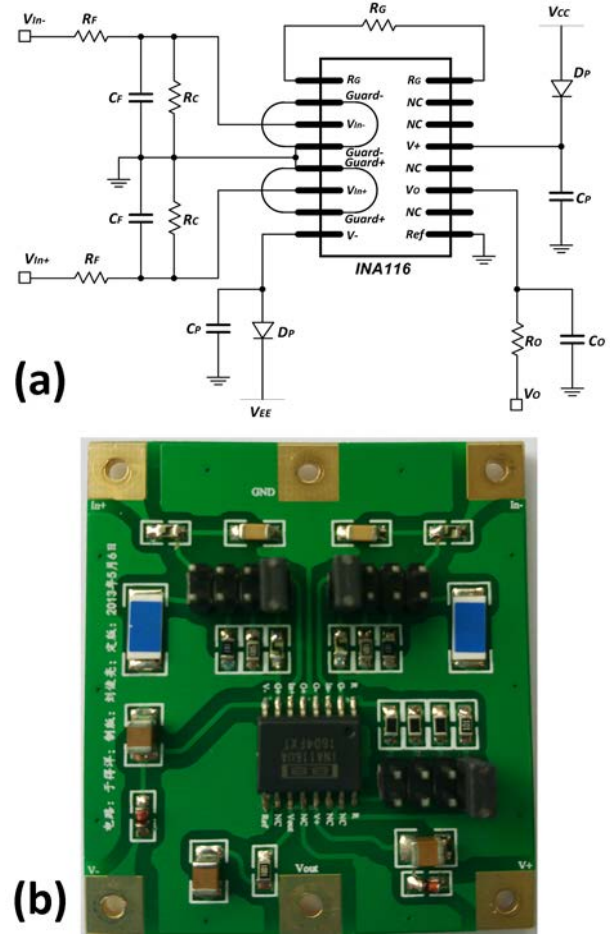

Fig. 1. (a) Circuit of the present current-to-voltage converter. (b) A photograph of the PCB implementation, i.e., the daughter board.

Sixteen daughter boards are mounted on a mother board, which is constructed on a 4-layer PCB, as shown in Fig. 2(a). In order to shunt leakage currents from the power supplies to the input stage wires, the non-inverting and the inverting input signal routes are arranged in independent layers, surrounded by guarding rings and sandwiched by guarding layers, as well as in cooperation with hollowing out parts, respectively. Two mother boards are mounted in a standard 1U 19-inch rack case, as shown in Fig. 2(b). It is found that microphonics caused by mechanical vibration [44] can produce detectable noise. Therefore, the cases are thickened, and the mother boards are firmly mounted with spring washers. In addition, all of the components are ultrasonic cleaned and sealed up by polyethylene film, in order to minimize leak current induced by dust accumulation. Benefit from the low power consuming of 24 mW of the INA116 chip, the temperature rise in the cases is negligible.

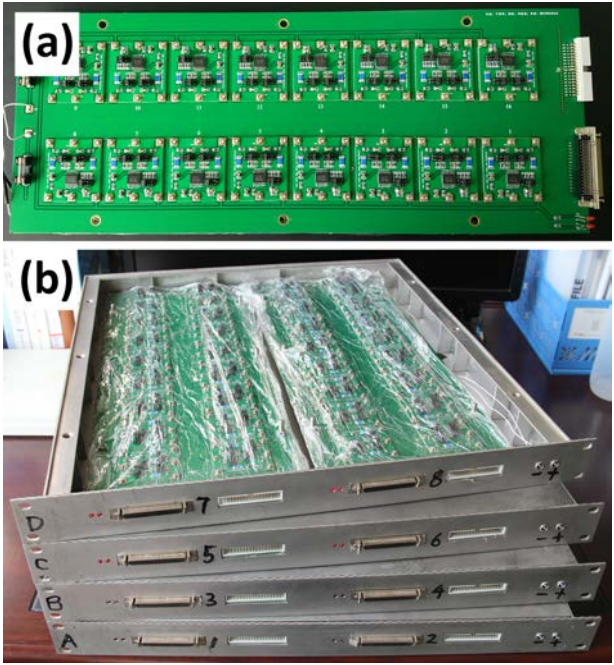

Fig. 2. (a) Photograph of the mother board which contains 16 daughter boards. (b) Four standard 1U 19-inch rack cases, each contains two mother boards inside.

The 128-channel output voltages are acquired by a computer equipped with 4 acquisition cards of National Instruments Model PCI-6224. [45] In order to save the acquisition channels, the referenced single-ended (RSE) mode rather than the differential mode [45] is employed. The acquisition program is written in LabVIEW, [46] which will be presented elsewhere in detail. In the present work, the sampling frequency is set to be 5 Hz in the program according to the signal bandwidth, to avoid the charge injection effect. [47]

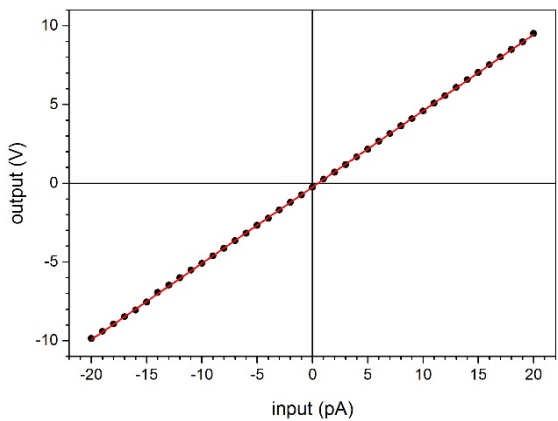

Fig. 3. Typical calibration of one channel of the picoammeter. In the calibration $R_C$ = 1 GΩ and $R_G$ = 100 Ω are selected, which corresponds to a gain of ~0.5 V/pA.

Every channel of the picoammeter are tested with a current source of Keithley Model 6221, and the zero-point offsets and the gains are calibrated, as shown in Fig. 3. The nonlinearity is negligible, and the zero-point offset is typically within several pA. The variation of gains among different channels is within ±5%, which is mainly caused by the inaccuracy of the 1 GΩ resistors $R_C$. The zero-point offsets and the gain adjustments are corrected in the acquisition program according to the calibrations.

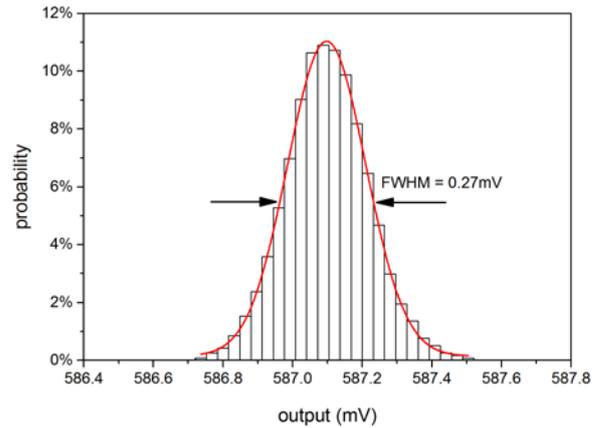

Fig. 4. Output amplitude statistical distribution of one of the best channels in a 4 hours test with a battery-based voltage source of 5.6 mV. The sampling rate is 5 Hz, and one column corresponds 32μV on the horizontal axis. During the test $R_C$ = 1 GΩ and $R_G$ = 511 Ω are selected, which corresponds to a gain of ~0.1 V/pA. The FWHM of the distribution is 0.27 mV, which corresponds to an input current resolution of ~2.7 fA.

In order to verify the system stability and resolution, a battery-based voltage source of 5.6 mV is applied on the non-inverting input of all channels in parallel. The source itself has advantages of long-term stability, extremely small interference and without ripples. Note that although $R_F$ has no effect on a constant current input, it does affect when a constant voltage source is employed. Hence, the current is different in different channels, depending on the inaccuracy of the resistor $R_F$ and $R_C$. The distribution of the output amplitude represents system resolution and stability. A typical output voltage statistical distribution for $R_C$ = 1 GΩ and $R_G$ = 511 Ω is shown in Fig. 4. During a 4-hour test with a 5 Hz sampling rate for each channel, it is found that the output amplitudes of the best channels satisfy quasi-Gaussian distributions with FWHMs smaller than 0.5 mV, which correspond to input current resolutions better than 5 fA. We have observed that the worst channels drift in the range of 20 mV, but the drift can be reduced by replacing the INA116 chips.

We emphasize that the input filter composed by $R_F$ and $C_F$ is a crucial stage in the present design, due to the RSE mode is employed. When a beam current of several picoamperes is measured, the induced 50 Hz perturbation on the input cable may deteriorate system resolution. A contrasting test is shown in Fig. 5. In which a current of 9.5 pA from the current source is feed into a channel of the picoammeter, while it is superposed with a 10 mV, 50 Hz perturbation through a capacitor of 1 nF. The capacitor $C_F$ = 0 pF, 10 pF, 100 pF and 1 nF are tested with $R_F$ = 1 GΩ, $R_C$ = 1 GΩ and $R_G$ = 100 Ω. It shows that $C_F$ = 1 nF and $R_F$ = 1 GΩ is sufficient to inhibit the 50 Hz interference from the AC power source.

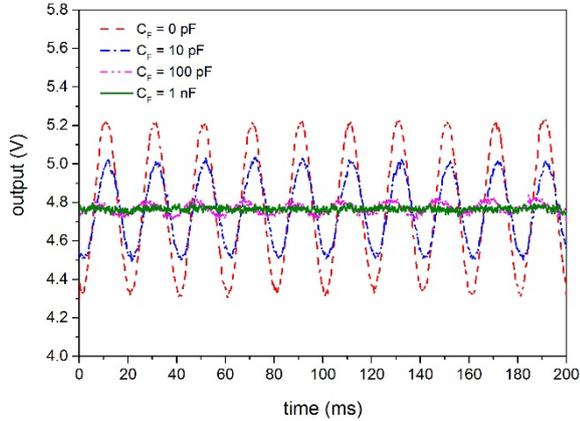

Fig. 5. Contrasting test of the input filter stage. A current of 9.5 pA is feed into a channel of the picoammeter, superposed with a 10 mV, 50 Hz perturbation through a capacitor of 1 nF. $R_C$ = 1 GΩ, $R_G$ = 100 Ω, $R_F$ = 1 GΩ while $C_F$ = 0 pF, 10pF, 100 pF and 1 nF are tested. Obtained by an oscilloscope of Model Tektronix TDS3032C which is extra bandwidth suppressed.

### III. THE 128-CHANNEL FARADAY CUP ARRAY

The 128-channel FCA is shown in Fig. 6(a). The 128 strip electrode array is fabricated on a double-side PCB, one version with electrode widths of 0.18 mm, lengths of 25 mm and gaps of 0.12 mm, while another with electrode widths of 0.3 mm, lengths of 25 mm and gaps of 0.2 mm, as shown in Fig. 6(b). The electrodes, which are guarded by a ground-ring, are employed to collect the beam current and therefore to obtain its one-dimensional distribution. A mesh is placed at 3 mm in front of the electrodes, which is connected to a negative bias voltage to suppress the secondary electrons. Another grounded mesh is placed at 3 mm in front of the suppressing mesh to confine its electric field. The meshes are made of 25 μm gold-coating tungsten wires, and arranged in arrays with a spatial period of 1 mm, with opening windows of 25×42 mm² in the first version and 25×68 mm² in the second version. In order to minimize the blurring of the spatial resolution by the suppressing fields applying to the beam, the wires are arranged perpendicular to the strip electrodes, which only deflect the beam parallel to the electrodes.

The kinetic energy distribution of secondary electrons emitted from the measurement electrodes depend on the incoming beam species and energy, but typically do not exceed several tens of eV with a distribution maximizing lower than 30 eV.[48, 49] The suppression of the secondary electrons is simulated by using the SIMION[50] software, as shown in Fig. 7. One hundred electrons are emitted with 50 eV from an electrode with a random angular distribution in the backward hemisphere. The bias voltage of the suppress mesh is −200 V. It shows that all the electrons are deflected back to the measurement electrodes, within a maximum range of ~4 mm. Note that most of the secondary electrons have much lower energies[47] and the therefore the diffusion range is actually much smaller. Space charge effects are not considered and expected to be negligibly small.

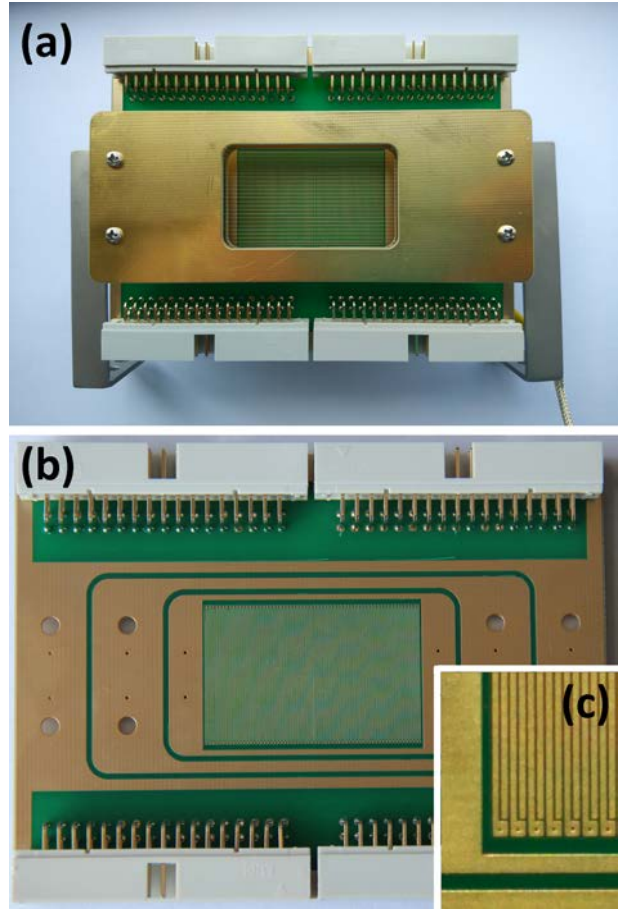

Fig. 6. Photograph of an FCA. (a) The overall view. (b) The PCB implementation of the 128-channel FCA by strip electrodes, which collect the charged particles. (c) Close-up view of the strip electrodes.

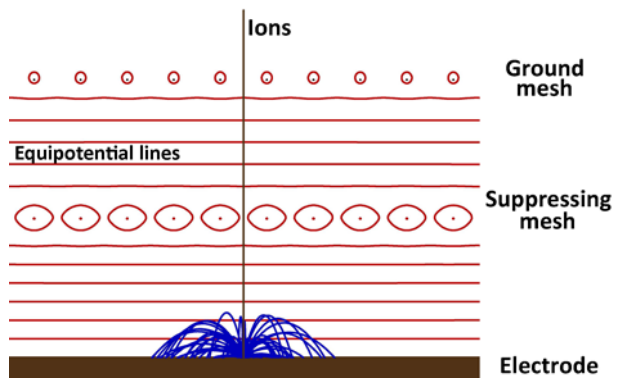

Fig. 7. Simulations trajectories of secondary electrons (blue lines) in the FCA. The secondary electrons are of 50 eV and emitted with a random angular distribution in the backward hemisphere in the graph, while the bias voltage of the suppressing mesh is −200V. Note that most of the secondary electrons have much lower energies and therefore the diffusion range is actually much smaller than this simulation.

## IV. RESULTS AND CONCLUSIONS

The 128 output signals of the FCA, as well as its ground, are routed out of the vacuum using three standard D-sub 50 UHV feed-through. The signals and the ground are connected to the non-inverting inputs and the inverting inputs of the picoammeter through ribbon cables, respectively. The ribbon cables are wrapped with aluminum foil to further reduce interference.

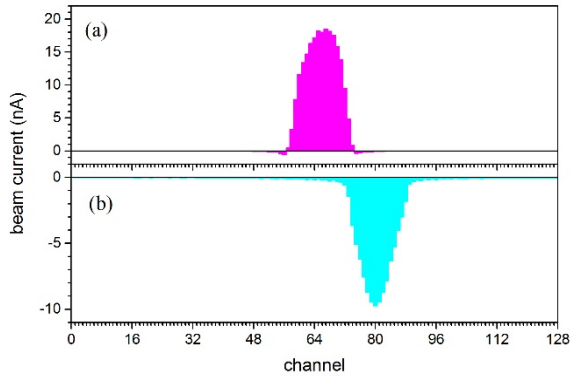

Fig. 8. Tests of the picoammeter system and the FCAs. (a) By a 60 keV $O^{3+}$ beam with the FCA of 0.18 mm electrodes and 0.12 mm gaps. (b) By a 2 keV electron beam with the FCA of 0.3 mm electrodes and 0.2 mm gaps. Exact boundaries of the ion beam can be obtained according to the negative signals produced by the diffusing secondary electrons. The width of the peaks stem from the collimating apertures of the beams.

As shown in Fig. 8(a) and Fig.8(b), the picoammeter system and the FCAs are tested by a 60 keV $O^{3+}$ beam (the FCA with 0.18 mm electrodes and 0.12 mm gaps) and by a 2 keV electron beam (the FCA with 0.3 mm electrodes and 0.2 mm gaps), respectively. It is interesting that the exact boundaries of the ion beam can be obtained, due to the diffusion of the secondary electrons which produce negative signals, as shown in Fig. 8(a). On the contrary, the boundaries of the electron beam are blurry, as shown in Fig. 8(b).

In conclusion, a 128-channel picoammeter system is constructed for measuring picoamperes continuous charged particle beam current distribution. Current resolution better than 5 fA is achieved by taking the advantages of high electric potential and narrow bandwidth in beam measurements. Two 128-channel FCAs are built, with opening areas of $25\times42$ mm$^2$ and $25\times68$ mm$^2$, and spatial resolutions of 0.3 mm and 0.5 mm. The system is tested by a 60 keV $O^{3+}$ beam and a 2 keV electron beam. It successfully identifies the exact boundaries when a positive charged beam is measured. The picoammeter system has been subjected to vigorous testing for more than 1000 hours using in total at different beam species, intensities and energies. Upon successful demonstration of the system, a two-dimensional fixed wires transmission beam profile monitor for continuous ion beam is under construction.

## ACKNOWLEDGMENT

We thank Weinian Ma and Ruishi Mao for their generous support on the acquisition cards. The ion beam test was carried out at the 320 kV platform for multidiscipline research with highly charged ions of Institute of Modern Physics, Chinese Academy of Sciences. This work is supported by the National Natural Science Foundation of China under Grant Nos. 11275240, U1332206 and 11205224.